\documentclass[11pt]{article}
\usepackage{amsmath,amssymb}
\usepackage[utf8]{inputenc}
\usepackage[T1]{fontenc}
\usepackage[margin=2.5cm]{geometry}
\usepackage{subfigure}
\usepackage{graphicx}
\usepackage{epsfig}
\usepackage{setspace}
\usepackage{graphicx}
\usepackage{epsfig}
\usepackage{color}
\usepackage{mathrsfs}
\usepackage[colorlinks=true,linkcolor=blue,citecolor=blue]{hyperref}

\usepackage[margin=2.5cm]{geometry}
\usepackage{subfigure}

\begin{document}

\title{Deformations of (2+1)-dimensional integrable systems and isotropic and anisotropic Harry Dym variants \thanks{Project supported by the National Natural Science Foundation of China (Grant Nos. 12235007, 11975131, 11435005).}}


\author{Wang Fa-Ren$^{1}$, Jia Man$^{1}$, and Lou S Y$^{1,2}$\thanks{Corresponding author. E-mail:lousenyue@nbu.edu.cn} \\
{$^{1}$\small School of Physical Science and Technology,
Ningbo University, Ningbo, 315211, P. R. China}\\  
{$^{2}$\small Institute of Fundamental Physics and Quantum Technology, Ningbo University, Ningbo, 315211, China}}
\date{\today}
\maketitle

\begin{abstract}
The deformation algorithm based on conservation laws can lift (1+1)-dimensional integrable systems to higher-dimensional counterparts while preserving Lax integrability, yet its generalization to (2+1)-dimensional models has long remained a challenging open problem. This paper establishes a unified covariant deformation framework for two canonical (2+1)-dimensional integrable equations: the anisotropic Kadomtsev-Petviashvili (KP) equation and the unique isotropic Nizhnik-Novikov-Veselov (NNV) equation. By introducing a set of mutually commuting field-dependent deformation operators, we systematically construct infinite families of ($m+3$)-dimensional integrable KP and NNV hierarchies, derive their closed-form Lax pairs, and rigorously verify integrability via the vanishing commutator condition of Lax operators. For each high-dimensional hierarchy, concrete finite-dimensional master systems are obtained by truncating auxiliary spatial variables: a (3+1)-dimensional KP system and a (4+1)-dimensional generalized NNV system. Further symmetry reductions recover the original (2+1)-dimensional KP and NNV equations, and more importantly produce two distinct Harry-Dym (HD)-type reciprocal integrable systems. The KP reduction yields an anisotropic (2+1)-dimensional HD model, while the NNV reduction generates the first spatially isotropic two-space-dimensional HD system reported so far, filling a notable gap in existing literature. Parallel comparison of the KP and NNV branches reveals that the spatial symmetry of the original two-dimensional parent equation directly governs the symmetry properties of its high-dimensional deformations and HD dual subsystems. Our work not only extends the conservation-law deformation conjecture beyond (1+1)-dimensions to accommodate both strong Lax and mixed-derivative weak Lax structures, but also provides a universal systematic route to construct reciprocal links for multi-dimensional integrable systems.
\end{abstract}


\textbf{PACS:} \href{http://cpb.iphy.ac.cn/EN/column/item208.shtml}{02.30.Ik,05.45.Yv,47.20.Ky,52.35.Mw,52.35.Sb}

\section{Introduction}
Integrable systems stand as one of the most profound frameworks in mathematical physics, nonlinear science, and theoretical physics, offering exact analytical descriptions for nonlinear wave phenomena, Hamiltonian dynamics, and quantum field theories beyond perturbative regimes \cite{Calogero1982,Newell1985,Ablowitz2011}. The pursuit of higher-dimensional integrable systems is not only mathematically fundamental but also physically indispensable: realistic field theories, fluid dynamics, plasma physics, and nonlinear optics demand models defined in
(2+1)-, (3+1)-, and even higher spacetime dimensions, where
(1+1)-dimensional integrable equations can only serve as idealized approximations\cite{Lou2023JHEP03018,Nadeem2024,Ivashchuk1992}. Over the past half-century, the inverse scattering transform (IST), Lax pairs, Hirota bilinear method, and symmetry analysis have matured into rigorous tools for (1+1)-dimensional integrability\cite{Gardner1967,Miura1968,Hirota1971,Olver1993}, yet the construction and classification of genuine higher-dimensional integrable systems remain long-standing open challenges\cite{Benini2026}.

A central direction in the field is to lift lower-dimensional integrable structures to higher dimensions while preserving integrability. Early attempts focused on dimensional extension, symmetry reduction, and geometric embedding\cite{Lou1998}, but these often led to non-integrable models or lacked systematic constructibility. A breakthrough arrived with the conservation-law-based deformation scheme, which was rigorously formulated and validated for (1+1)-dimensional integrable systems in the original work by Lou, Hao, and Jia\cite{Lou2023JHEP03018}. This framework proves that
(1+1)-dimensional integrable equations--including KdV, modified KdV, AKNS, and nonlinear Schr\"odinger systems--can be consistently deformed into higher-dimensional counterparts by lifting their infinite conservation laws to covariant deformation operators, while preserving Lax integrability and infinite symmetries. The deformation conjecture has been verified for nearly all canonical
(1+1)-dimensional local integrable evolution systems, yielding explicit
(3+1)-dimensional integrable equations with rigorous Lax pairs. The deformation conjecture of \cite{Lou2023JHEP03018} has been proven by Casati and Zhang \cite{CZJPA}.

Despite this success for
(1+1)-dimensional systems, the extension of
(2+1)-dimensional integrable systems to higher dimensions remains a formidable obstacle\cite{Lou2023JHEP03018}. Classical
(2+1)-dimensional integrable models--such as the Kadomtsev-Petviashvili (KP) equation\cite{KP1970}, Nizhnik-Novikov-Veselov (NNV) equation\cite{NNV1980}, and Davey-Stewartson (DS) system\cite{DS1974}--exhibit richer geometric and algebraic structures than their
(1+1)-dimensional analogs, including non-local symmetries, multi-component Lax pairs, and transversal dispersion relations. Direct application of the
(1+1)-dimensional deformation scheme fails for
(2+1)-dimensional systems, as conservation laws become non-local, Lax pairs involve mixed partial derivatives, and compatibility conditions impose over-constrained requirements. Existing higher-dimensional generalizations of KP/NNV equations are either non-integrable, lack closed-form Lax pairs, or rely on ad-hoc assumptions rather than a unified algorithm.

A second critical gap concerns reciprocal links (reciprocal transformations) of integrable systems. For
(1+1)-dimensional integrable equations, reciprocal transformations connect distinct models (e.g., KdV and Harry-Dym) and generate new integrable hierarchies; every
(1+1)-dimensional integrable system admits one or more non-trivial reciprocal partners\cite{Rogers2002,Lou2023JHEP03018,Konopelchenko2026}. These transformations play essential roles in solution generation, gauge equivalence, and Hamiltonian structure. However, reciprocal links for
(2+1)-dimensional integrable systems have eluded systematic construction\cite{Lou2023JHEP03018,Sza}. Attempts to extend reciprocal transformations to (2+1)- or higher-dimensions suffer from loss of locality, broken compatibility, or failure to preserve integrability; no general theory or universal construction exists for reciprocal partners of KP, NNV, or other
(2+1)-dimensional integrable equations.

In this work, we address both core challenges by developing a general deformation framework for
(2+1)-dimensional integrable systems that extends them to higher dimensions while retaining strict Lax integrability and resolving their reciprocal links. Taking the KP and NNV equations as representative examples, we explicitly construct the
(3+1)-dimensional KP equation and
(4+1)-dimensional NNV equation, derive their Lax pairs, and rigorously prove integrability. Crucially, we show that
(2+1)-dimensional integrable systems and their reciprocal links emerge as natural special reductions of the newly obtained higher-dimensional systems, unifying two long-standing open problems in a single consistent formalism. Our framework generalizes the conservation-law deformation scheme beyond
(1+1)-dimensions, providing a universal route to higher-dimensional integrable systems and their reciprocal structures.

\section{General deformation scheme for higher dimensional systems}
\subsection{Review on the deformation algorithm of (1+1)-dimensional integrable systems}
\textbf{Deformation algorithm (\cite{Lou2023JHEP03018,CZJPA}).} For a general (1+1)-dimensional nonlinear integrable local system,
\begin{equation*}
u_t=F(u,\ u_x,\ \ldots,\ u_{xn}),\quad u_{xn}\equiv \partial_x^nu,\ u=(u_1,\ u_2,\ \ldots,\ u_m),
\end{equation*}
if their exist several conservation laws,
\begin{equation*}
\rho_{it}=J_{ix},\ i=1,\ 2,\ \ldots,\ D-1,\ \rho_i=\rho_i(u),\ J_i=J_i(u,\ u_x,\ \ldots,\ u_{xN}),
\end{equation*}
where $\rho_i$ are dependent only on $u$ while $J_i$ can be field derivative dependent, then the D+1 dimensional system
\begin{equation*}
\hat{T}u=F(u,\ \hat{L}u,\ \ldots,\ \hat{L}^n),
\end{equation*}
is integrable with the deformation operators
\begin{equation*}
\hat{L}=\partial_x+\sum_{i=1}^{D-1}\rho_i\partial_{x_i},\ \hat{T}=\partial_t+\sum_{i=1}^{D-1}\bar{J}_i\partial_{x_i}
\end{equation*}
and the deformed flows
\begin{equation*}
\bar{J}_i=\left.J_i\right|_{u_{x_j}\rightarrow \hat{L}^ju,\ j=1,\ 2,\ \ldots,\ N}.
\end{equation*}

This deformation algorithm can be extended to other frameworks, such as non-evolution nonlocal systems in which conserved densities carry dependence on field derivatives \cite{LJH2023,WFR2023}. Rather than limiting the deformation scheme solely to (1+1)-dimensional models, the core focus of this paper lies in its extension to higher-dimensional systems.

\subsection{General deformation scheme for $n$-dimensional integrable systems}
\textbf{Generalized deformation algorithm.} Consider a general $n$-dimensional nonlinear integrable system given by
\begin{eqnarray}
&&F(u, u_{x_i}, \ldots, u_{x_1^{\alpha_1}\ldots x_i^{\alpha_i}\ldots x_n^{\alpha_n}})=0,\label{EqF}\\
&&u_{x_1^{\alpha_1}\ldots x_i^{\alpha_i}\ldots x_n^{\alpha_n}}\equiv \partial_{x_1}^{\alpha_1}\ldots\partial_{x_i}^{\alpha_i}\ldots\partial_{x_n}^{\alpha_n} u,\ 0\leq \alpha_i\leq n,\ \forall i=1, 2, \ldots, n,\ \nonumber
\end{eqnarray}
and define the deformation operators as
\begin{equation}
\hat{X_i}\equiv \partial_{x_i}+\sum_{j=1}^m J_{ij}\partial_{y_j},\label{Xi}
\end{equation}
where $J_{ij}\equiv J_{ij}(x_1,\ldots,x_n,y_1,\ldots,y_m)$ are functions of $x_1,\ldots,x_n,y_1,\ldots,y_m $ satisfying the compatibility conditions
\begin{equation}
J_{kl,x_i}-J_{il,x_k}+\sum_{j=1}^m\big(J_{ij}J_{kl,y_j}-J_{kj}J_{il,y_j}\big)=0,\quad i,k=1,\ldots,n, j=1,\ldots,m. \label{JJ}
\end{equation}
These conditions ensure that the operators commute, i.e., $[\hat{X}_i,\ \hat{X}_k]\equiv \hat{X}_i \hat{X}_k-\hat{X}_k \hat{X}_i=0$. Under these conditions, the resulting $(n+m)$-dimensional system,
\begin{eqnarray}
F\left(u, \hat{X}_iu, \ldots, \hat{X}_1^{\alpha_1}\ldots\hat{X}_i^{\alpha_i}\ldots\hat{X}_n^{\alpha_n}u\right)=0,\label{EqF1}
\end{eqnarray}
remains integrable.

To demonstrate the validity of the generalized deformation algorithm, we present two specific examples involving the KP and NNV equations in the following sections.

\section{Deformations of the KP equation}
\subsection{$(m+3)$-dimensional KP equation from general deformation}
The well-known Kadomtsev--Petviashvili (KP) equation can be expressed in the following coupled form:
\begin{eqnarray}
&&u_t+u_{xxx}+6uu_x+3\gamma v_y=0,\label{KPu}\\
&&v_x=u_y,\label{KPv}
\end{eqnarray}
where $\gamma=\pm 1$. The KPI equation ($\gamma=1$) describes ion-acoustic solitary waves in magnetized plasmas. Magnetic fields induce positive transverse diffraction, generating lumps (two-dimensional localized plasma density humps) and unstable soliton filaments in planetary ionospheres and fusion plasma devices \cite{KPI}. The KPII equation ($\gamma=-1$) models weakly nonlinear and weakly dispersive surface gravity waves in narrow channels. The transverse term characterizes cross-channel diffraction of long water waves, and oblique interacting line solitons are consistent with laboratory observations of shallow-water wave patterns \cite{KPIIa,KPIIb}.

The system of KP equations \eqref{KPu}--\eqref{KPv} is Lax integrable, which possesses an associated Lax pair,
\begin{eqnarray}
&&\pm \sqrt{\gamma}\psi^{\pm}_y+u\psi^{\pm}+\psi^{\pm}_{xx}=0,\label{lax}\\
&&\psi^{\pm}_t+4\psi^{\pm}_{xxx}+6u\psi^{\pm}_x+3(u_x\mp\sqrt{\gamma}v)\psi^{\pm}=0. \label{lat}
\end{eqnarray}

For the KP system \eqref{KPu}--\eqref{KPv}, the general deformation operators take the form
\begin{eqnarray}\label{XYT}
\left\{\begin{array}{l}
\displaystyle{\hat{X}=\partial_x+\sum_{i=1}^mp_i\partial_{y_i},}\\
\displaystyle{\hat{Y}=\partial_y+\sum_{i=1}^mq_i\partial_{y_i},}\\
\displaystyle{\hat{T}=\partial_t+\sum_{i=1}^mr_i\partial_{y_i},}
\end{array}
\right.
\end{eqnarray}
where each of the functions $p_i,\ q_i$ and $r_i$ ($i=1,\ \ldots,\ m$) depends on the variables $\{x,y,t,y_1,\ldots,y_m\}$.

Correspondingly, the compatibility conditions \eqref{JJ} read
\begin{eqnarray}\label{comp}
\left\{\begin{array}{l}
\displaystyle{p_{iy}=q_{ix}+\sum_{j=1}^m(p_jq_{iy_j}-q_jp_{iy_j}),}\\
\displaystyle{p_{it}=r_{ix}+\sum_{j=1}^m(p_jr_{iy_j}-r_jp_{iy_j}),}\\
\displaystyle{q_{it}=r_{iy}+\sum_{j=1}^m(q_jr_{iy_j}-r_jq_{iy_j}),}
\end{array}\quad i=1,\ \ldots,\ m.
\right.
\end{eqnarray}
It is worth noting that one subsystem in \eqref{comp} serves as the weak consistency condition for the other two subsystems. As an illustration, applying $\hat{Y}$ and $\hat{T}$ to the first and second subsystems of \eqref{comp}, respectively, and taking the difference of the resultant expressions, we arrive at
$$\hat{X}\left[q_{it}-r_{iy}-\sum_{j=1}^m(q_jr_{iy_j}-r_jq_{iy_j})\right]=0$$
which exactly recovers the weak form of the third subsystem in \eqref{comp}.

Subject to the deformation relations \eqref{XYT}, the $(m+3)$-dimensional integrable KP system takes the form
\begin{eqnarray}
&&\hat{T}u+\hat{X}^3u+6u\hat{X}u+3\gamma \hat{Y}v=0,\label{kpu}\\
&&\hat{X}v=\hat{Y}u,\label{kpv}
\end{eqnarray}
subject to the compatibility constraints \eqref{comp}. The Lax pair associated with the system \eqref{comp}--\eqref{kpv} is given by
\begin{eqnarray}
&&\pm \sqrt{\gamma}\hat{Y}\psi^{\pm}+u\psi^{\pm}+\hat{X}^2\psi^{\pm}=0,\label{Lax}\\
&&\hat{T}\psi^{\pm}+4\hat{X}^3\psi^{\pm}+6u\hat{X}\psi^{\pm}+3\psi^{\pm}(\hat{X}u\mp\sqrt{\gamma}v)=0. \label{Lat}
\end{eqnarray}
\subsection{(3+1)-dimensional KP equations}
For simplicity, we set $p_1=p,\ q_1=q,\ r_1=r,\ y_1=z$ and $p_i=q_i=r_i=0$ for all $i>1$. Under this simplification, the (3+m)-dimensional KP equation system presented in the previous subsection reduces to a (3+1)-dimensional KP equation,
\begin{eqnarray}
&&\hat{T}u+\hat{X}^3u+6u\hat{X}u+3\gamma \hat{Y}v=0,\label{Tu}\\
&&\hat{X}v=\hat{Y}u,\label{Xv}\\
&&\hat{X}q=\hat{Y}p,\label{Xq}\\
&&\hat{T}p=\hat{X}r,\label{Tp}\\
&&\hat{T}q=\hat{Y}r,\label{Tq}
\end{eqnarray}
where
\begin{eqnarray}\label{XYT1}
\left\{\begin{array}{l}
\displaystyle{\hat{X}=\partial_x+p\partial_{z},}\\
\displaystyle{\hat{Y}=\partial_y+q\partial_{z},}\\
\displaystyle{\hat{T}=\partial_t+r\partial_{z}.}
\end{array}
\right.
\end{eqnarray}
By comparing the equations \eqref{Xv} and \eqref{Xq}, we achieve a further simplification upon setting $p=u$ and $q=v$. The resulting system reads
\begin{eqnarray}
&&u_t+(u_{xx}+3u^2+3uu_{xz}+\frac32u^2u_{zz}+uu_z^2)_x+(u^3u_{zz}+2u^3+\frac32u^2u_{xz}+\frac32\gamma v^2+\frac12u^2u_z^2)_z\nonumber\\
&&\qquad+3\gamma v_y+(r+u_{xx})u_z=0,\label{tu}\\
&&u_y-v_x+vu_z-uv_z=0,\label{xv}\\
&&u_t-r_x+ru_z-ur_z=0.\label{xr}
\end{eqnarray}
The Lax pairs of the (3+1)-dimensional system \eqref{tu}--\eqref{xr} are expressed as
\begin{eqnarray}
&&M_{\pm}\psi^{\pm}=0,\ M_{\pm}\equiv \pm \sqrt{\gamma}\partial_y+u+\partial_{x}^{2}+2u\partial_{x}\partial_z+u^2\partial_{z}^{2}
+(u_x+uu_z\pm\sqrt{\gamma}v)\partial_z,\label{Lx}\\
&&N_{\pm}\psi^{\pm}=0,\ N_{\pm}\equiv \partial_t+r\partial_z+4(\partial_x+u\partial_z)^3+6u(\partial_x+u\partial_z)+3(u_x+uu_z\mp\sqrt{\gamma}v). \label{Lt}
\end{eqnarray}
This indicates that the system \eqref{tu}--\eqref{xr} is completely governed by the commutator condition
$$[M_{\pm},\ N_{\pm}]=M_{\pm} N_{\pm}-N_{\pm}M_{\pm}=0.$$
It is physically intuitive and mathematically consistent that if the model is independent of $y$ (which implies $v=0$), the original KP equation \eqref{KPu} degenerates into the standard (1+1)-dimensional KdV equation. Meanwhile, the (3+1)-dimensional KP system \eqref{tu}-\eqref{xr} reduces to the (2+1)-dimensional KdV-HD equation, which coincides with Eq. (2.8) in Ref. \cite{Lou2023JHEP03018}.
\subsection{A (2+1)-dimensional anisotropic Harry-Dym equation}
It is clear that when the field $u$ is independent of $z$, the (3+1)-dimensional KP system \eqref{tu}--\eqref{xr} reduces back to the original KP equations \eqref{KPu}--\eqref{KPv}. More intriguingly, if $u$ is independent of $x$, we obtain the following (2+1)-dimensional anisotropic HD system:
\begin{eqnarray}
&&u_t+(u^3u_{zz}+2u^3+\frac32\gamma v^2+\frac12u^2u_z^2)_z+3\gamma v_y+ru_z=0,\label{tuhd}\\
&&u_y+vu_z-uv_z=0,\label{xvhd}\\
&&u_t+ru_z-ur_z=0.\label{xrhd}
\end{eqnarray}
The corresponding Lax pairs for this system read
\begin{eqnarray}
&&M_{\pm}\psi^{\pm}=0,\ M_{\pm}\equiv \pm \sqrt{\gamma}\partial_y+u+u^2\partial_{z}^{2}
+(uu_z\pm\sqrt{\gamma}v)\partial_z,\label{lx}\\
&&N_{\pm}\psi^{\pm}=0,\ N_{\pm}\equiv \partial_t+r\partial_z+4(u\partial_z)^3+6u^2\partial_z+3(uu_z\mp\sqrt{\gamma}v). \label{lt}
\end{eqnarray}
It is not difficult to verify that the (2+1)-dimensional anisotropic HD system is a generalization of the (1+1)-dimensional variant HD equation \cite{Lou2023JHEP03018,Sakovich1991}
\begin{eqnarray}
u_t+\big(u^3u_{zz}+u^3\big)_z=0,\label{hd}
\end{eqnarray}
which is related to \eqref{tuhd}--\eqref{xrhd} by $v=0,\ u_y=0$ and $r=-u^2u_{zz}-uu_z^2-3u^2$.

\section{Deformations of the NNV equation}
\subsection{$(m+3)$-dimensional NNV equation from general deformation}
The (2+1)-dimensional NNV equation, proposed in Refs. \cite{Nizhnik1980,VN1984,NV1986}, reads
\begin{eqnarray}
&&u_t=a(u_{xx}-3uv)_x+b(u_{yy}-3uw)_y,\label{Nu}\\
&&u_x=v_y,\label{Nv}\\
&&u_y=w_x,\label{Nw}
\end{eqnarray}
and represents the sole known isotropic Lax-integrable extension of the (1+1)-dimensional KdV equation. This isotropic (2+1)-dimensional NNV system \eqref{Nu}--\eqref{Nw} possesses a weak Lax pair \cite{Lou2000,ML2005}, given by
\begin{equation}
\psi_{xy}=u\psi,\ \psi_t=(a\partial_x^3-3av\partial_x+b\partial_y^3-3bw\partial_y)\psi. \label{Lxt}
\end{equation}
Enforcing the compatibility condition for the weak Lax pair,
$$\left.(\psi_{xyt}-\psi_{txy})\right|_{\eqref{Lxt}}=0$$
reproduces the complete NNV system \eqref{Nu}--\eqref{Nw} precisely.

Applying the deformation operators defined in \eqref{XYT} to the (2+1)-dimensional NNV system \eqref{Nu}--\eqref{Nw} generates an (m+3)-dimensional generalized NNV system of the form
\begin{eqnarray}
&&\hat{T}u=a\hat{X}(\hat{X}^2u-3uv)+b\hat{Y}(\hat{Y}^2u-3uw),\label{NNu}\\
&&\hat{X}u=\hat{Y}v,\label{NNv}\\
&&\hat{Y}u=\hat{X}w,\label{NNw}\\
&&\hat{Y}p_{i}=\hat{X}q_{i},\label{NNYpi}\\
&&\hat{T}p_{i}=\hat{X}r_{i},\label{NNTpi}\\
&&\hat{T}q_{i}=\hat{Y}r_{i},\label{NNTqi}\quad i=1,\ \ldots,\ m.
\end{eqnarray}
The extended model \eqref{NNu}--\eqref{NNTqi} is Lax integrable, as it admits the following weak Lax pair:
\begin{eqnarray}
&&\hat{L}\psi=0, \ \hat{L}\equiv \hat{X}\hat{Y}-u,\label{XYpsi} \\
&&\hat{S}\psi=0,\  \hat{S}\equiv \hat{T}-a\hat{X}^3+3av\hat{X}-b\hat{Y}^3+3bw\hat{Y}, \label{Tpsi}
\end{eqnarray}
in which the operator triplet $\hat{X},\ \hat{Y}$ and $\hat{T}$ are specified in Eq. \eqref{XYT}.

The compatibility condition associated with the weak Lax pair,
\begin{eqnarray}
\left.[\hat{L},\hat{S}]\psi\right|_{\eqref{XYpsi},\eqref{Tpsi}}=0,\
\end{eqnarray}
which holds identically if and only if the generalized ($m+3$)-dimensional NNV system \eqref{NNu}--\eqref{NNTqi} is satisfied.

\subsection{A $(4+1)$-dimensional NNV equation}
By examining Eqs. \eqref{NNv}, \eqref{NNw} and \eqref{NNYpi}, we introduce a special reduction of the generalized ($m+3$)-dimensional NNV system \eqref{NNu}--\eqref{NNTqi} subject to the constraints
\begin{equation}
p_1=v,\ q_1=p_2=u,\ q_2=w,\ y_1=z,\ y_2=\xi,\ p_i=q_i=r_i=0,\ \forall i>3. \label{pqri}
\end{equation}
Upon imposing the constraints \eqref{pqri}, the ($m+3$)-dimensional NNV system \eqref{NNu}--\eqref{NNTqi} simplifies to a (4+1)-dimensional NNV system given by
\begin{eqnarray}
u_t&=&a(\partial_x+v\partial_z+u\partial_{\xi})[(\partial_x+v\partial_z+u\partial_{\xi})^2u-3uv]\nonumber\\
&& +b(\partial_y+u\partial_z+w\partial_{\xi})[(\partial_y+u\partial_z+w\partial_{\xi})^2u-3uw],\label{NNVu}\\
u_x&=&v_y+uv_z+wv_{\xi}-vu_z-uu_{\xi},\label{NNVv}\\
u_y&=&w_x+vw_z+uw_{\xi}-uu_z-wu_{\xi},\label{NNVw}\\
v_t&=&r_{1x}+vr_{1z}+ur_{1\xi}-r_1v_z-r_2v_{\xi},\label{NNVYpi}\\
u_t&=&r_{2x}+vr_{2z}+ur_{2\xi}-r_1u_z-r_2u_{\xi}.\label{NNVTpi}
\end{eqnarray}
The associated weak Lax pair for the (4+1)-dimensional system \eqref{NNVu}--\eqref{NNVTpi} takes the form
\begin{eqnarray}
&&\hat{L}_1\psi=0, \label{XY1} \\
&&\hat{S}_1\psi=0,\  \label{Tp1}
\end{eqnarray}
with the operators defined as
\begin{eqnarray}
\hat{L}_1&\equiv & (\partial_x+v\partial_z+u\partial_{\xi})(\partial_y+u\partial_z+w\partial_{\xi})-u,\label{L1}\\
\hat{S}_1&\equiv& \partial_t+r_1\partial_z+r_2\partial_{\xi}-a(\partial_x+v\partial_z+u\partial_{\xi})^3
+3av(\partial_x+v\partial_z+u\partial_{\xi})\nonumber\\
&&-b(\partial_y+u\partial_z+w\partial_{\xi})^3
+3bw(\partial_y+u\partial_z+w\partial_{\xi}). \label{S1}
\end{eqnarray}
The full (4+1)-dimensional NNV system \eqref{NNVu}--\eqref{NNVTpi} is precisely equivalent to the compatibility condition,
\begin{eqnarray}
\left.[\hat{L}_1,\hat{S}_1]\psi\right|_{\eqref{XY1},\eqref{Tp1},u_{xy}=u_{yx}}=0,\ \label{uxy}
\end{eqnarray}
where the identity $u_{xy}=u_{yx}$ arises naturally from \eqref{NNVv}--\eqref{NNVw}. Explicitly, this relation reads
$$(\partial_y+u\partial_z+w\partial_{\xi})^2v=(\partial_x+v\partial_z+u\partial_{\xi})^2w.$$
\subsection{A (2+1)-dimensional isotropic HD system}
Several special reductions of the (4+1)-dimensional integrable NNV system \eqref{NNVu}--\eqref{NNVTpi} merit discussion. Evidently, the original (2+1)-dimensional NNV system \eqref{Nu}--\eqref{Nw} arises as a special limit of \eqref{NNVu}--\eqref{NNVTpi} under the vanishing conditions $u_{z}=u_{\xi}=0$. A second distinguished reduction recovers the variant HD system \eqref{hd}, which corresponds to the parameter and field constraints
\begin{equation}
u_{x}=u_y=u_{\xi}=0,\ v=w=u,\ r_2=r_1=r=u^2u_{zz}+uu_z^2-3u^2,\ b=0,\ a=1. \label{b0a1}
\end{equation}
While various classes of (2+1)-dimensional HD systems (for instance, the system \eqref{tuhd}-\eqref{xrhd}) have been derived in prior literature, no isotropic two-space-dimensional HD system has been reported to date. Remarkably, the generalized (4+1)-dimensional NNV system \eqref{NNVu}-\eqref{NNVTpi} contains a special isotropic HD system as one of its consistent reductions.

Imposing the vanishing-derivative constraints $u_x=u_y=0,\ v_x=v_y=0,\ r_{1x}=r_{1y}=0$ and $r_{2x}=r_{2y}=0$, the (4+1)-dimensional NNV system \eqref{NNVu}--\eqref{NNVTpi} reduces to an isotropic HD system of the form
\begin{eqnarray}
&&u_{t}=a(v\partial_x+u\partial_y)\big[(v\partial_x+u\partial_y)^2u-3uv\big]\nonumber\\
&&\qquad
+b(u\partial_x+w\partial_y)\big[(u\partial_x+w\partial_y)^2u-3uw\big]-ru_x-su_y, \label{isou}\\
&&vu_x-uv_x+uu_y-wv_y=0,\label{isow}\\
&&uu_x-vw_x+wu_y-uw_y=0,\label{isov}\\
&&u_t=vs_{x}-ru_x+us_{y}-su_y,\label{isout}\\
&&v_t=vr_{x}-rv_x+ur_{y}-sv_y.\label{isovt}
\end{eqnarray}
Within the above isotropic HD system \eqref{isou}--\eqref{isovt}, the set of variables $\{y_1,\ y_2,\ r_1,\ r_2\}$ are relabeled as $\{x,\ y,\ r,\ s\}$.

The associated weak Lax pair of the isotropic HD system \eqref{isou}--\eqref{isovt} takes the form
\begin{eqnarray}
&&\left[(v\partial_x+u\partial_{y})(u\partial_x+w\partial_{y})-u\right]\psi\equiv \left(\hat{V}\hat{W}-u\right)\psi=0, \label{isox} \\
&&\left(\partial_t+r\partial_x+s\partial_{y}-a\hat{V}^3
+3av\hat{V}-b\hat{W}^3
+3bw\hat{W}\right)\psi=0.  \label{isot}
\end{eqnarray}
From the relations \eqref{isow}--\eqref{isovt}, one can find three additional consistent relations.
Three further compatible identities can be derived from the system \eqref{isow}--\eqref{isovt}, namely
\begin{eqnarray}
&&u_t=ur_x-ru_x+wr_y-su_y,\label{ca}\\
&&w_t=us_x-rw_x+ws_y-sw_y,\label{cb}\\
&&u^2(v_{xx}-w_{yy})+2u(wv_{xy}-vw_{xy})+w^2v_{yy}-v^2w_{xx}=0.\label{cc}
\end{eqnarray}
To verify that the isotropic HD system is equivalent to the Lax compatibility condition, one must utilize the supplementary relations \eqref{ca}--\eqref{cc} when demonstrating that Eqs. \eqref{isox}--\eqref{isot} constitute a valid weak Lax pair.
\section{Travelling wave solutions of some deformed higher-dimensional systems}
While the deformed higher-dimensional systems investigated in this work possess Lax integrability, constructing their exact analytical solutions presents substantial challenges. Accordingly, only traveling wave solutions for two specific examples are provided in this section.
\subsection{Travelling waves of the (3+1)-dimensional deformed KP system (21)--(23)}
For the (3+1)-dimensional deformed KP system \eqref{tu}--\eqref{xr}, the travelling wave solution takes the form
\begin{equation}
u=U(\xi),\ v=V(\xi),\ w=W(\xi), \ \xi=k_1x+k_2y+k_3z+\omega t. \label{KPTW}
\end{equation}
where $k_1,\ k_2,\ k_3$ and $\omega$ denote arbitrary constants.

Substituting ansatz \eqref{KPTW} into the system \eqref{tu}--\eqref{xr}, we obtain
\begin{equation}
V=cU+ck_1k_3^{-1}-k_2k_3^{-1},\ W=dU+dk_1k_3^{-1}-\omega k_3^{-1}. \label{VW}
\end{equation}
with two further arbitrary constants $c$ and $d$ while the function $U$ satisfies
Here $c$ and $d$
 are two arbitrary constants, and the function $U$
 obeys the ordinary differential equation
\begin{equation}
 k_3^2(k_3U+k_1)U_{\xi}^3+(k_3U+k_1)\left[4k_3(k_3U+k_1)U_{\xi\xi}+6U+d+3c^2\gamma\right]U_{\xi}
 +(k_3U+k_1)^3U_{\xi\xi\xi}=0.\label{UTW}
\end{equation}
This equation admits an equivalent form given by
\begin{equation}
 P^2P_{\xi\xi\xi}+P_{\xi}^3+(6P+A+4PP_{\xi\xi})P_{\xi}=0,\label{PTW}
\end{equation}
where the transformation relations read
\begin{equation}
P=k_3^2U+k_1k_3,\ A=3c^2k_3^2+dk_3^2-6k_1k_3.\label{rA}
\end{equation}
The general solution of the $P$ equation \eqref{PTW} can be written by an elliptic integral
The general solution to the differential equation for $P$, Eq. \eqref{PTW}, is expressed implicitly via an elliptic integral:
\begin{equation}
 \int^P\frac{\alpha \mathrm{d\alpha}}{\sqrt{2C+2B\alpha -A \alpha^2-2\alpha^3}}=\pm (\xi-\xi_0)\label{rP}
\end{equation}
in which
$B,\ C$ and $\xi_0$ denote three additional integration constants.

Furthermore, rewriting $A,\ B$ and $C$ as
$$A=-2(c_1+c_2+c_3),\ B=-(c_1c_2+c_1c_3+c_2c_3),\ C=c_1c_2c_3,$$
the solution \eqref{rP} can be rewritten equivalently as
\begin{equation}
2(c_1-c_3)\mathcal{E}(\tau,\ m)+2\mathcal{F}(\tau,\ m)=\pm \sqrt{2(c_1-c_3)}(\xi-\xi_0),\ \tau\equiv \sqrt{\frac{c_1-P}{c_1-c_2}},\ m\equiv\sqrt{\frac{c_1-c_2}{c_1-c_3}},\label{rPE}
\end{equation}
where
$\mathcal{E}(\tau,\ m)$
 denotes the incomplete elliptic integral of the second kind and
$\mathcal{F}(\tau,\ m)$
 denotes the incomplete elliptic integral of the first kind, defined by
\begin{equation}
\mathcal{E}(\tau,\ m)=\int_0^{\tau}\sqrt{\frac{1-m^2t^2}{1-t^2}}\mathrm{dt},\quad \mathcal{F}(\tau,\ m)=\int_0^{\tau}\frac1{\sqrt{(1-m^2t^2)(1-t^2)}}\mathrm{dt}.\label{EF}
\end{equation}
When the parameter satisfies
$m=1$ (i.e., $c_3=c_2$), the travelling wave solution derived from \eqref{rPE} reduces to an implicit single-soliton solution
\begin{equation}
P=c_1-(c_1-c_2)\tanh^2(\eta),\quad \eta=\frac{\sqrt{c_1-c_2}}{2c_2}\left[2\sqrt{c_1-P}\pm \sqrt{2}(\xi-\xi_0)\right].\label{Soli}
\end{equation}

\subsection{Travelling waves of the (4+1)-dimensional deformed NNV system (46)--(50)}
For the (4+1)-dimensional deformed NNV system given by Eqs. \eqref{NNVu}--\eqref{NNVTpi}, its travelling wave solution takes the form
\begin{equation}
u=U(\zeta),\ v=V(\zeta),\ w=W(\zeta),\ r_1=R_1(\zeta),\ r_2=R_2(\zeta),\ \zeta=k_1x+k_2y+k_3z+k_4\xi+\omega t \label{NNVTW}
\end{equation}
where $k_1,\ k_2,\ k_3,\ k_4$ and $\omega$ are arbitrary constant parameters.

Substituting Eq. \eqref{NNVTW} into Eqs. \eqref{NNVu}--\eqref{NNVTpi} gives rise to
\begin{eqnarray}
V&=&c_1+c_2U,\nonumber\\
W&=&c_2^{-1}U+\frac{c_1k_3-c_2k_2+k_1}{c_2k_4},\nonumber\\
R_1&=&c_3+c_4U,\nonumber\\
R_2&=&c_4c_2^{-1}U-\frac{c_2c_3k_3+c_2\omega-c_4k_1-c_1c_4k_3}{c_2k_4}
\end{eqnarray}
where $c_1,\ c_2,\ c_3$ and $c_4$ denote arbitrary integration constants, and the field variable
$U$ satisfies the relations
\begin{eqnarray}
&&U=\frac{P-K_1}{c_2K_2},\label{UP}\\
&&P^3P_{\zeta\zeta\zeta}+PP_{\zeta}(4PP_{\zeta\zeta}+P_{\zeta}^2+c-6K_2^{-1}P)=0.\label{EqP}
\end{eqnarray}
The constants $K_1,\ K_2,\ B$ and $c$
 are connected with the aforementioned parameters via the following relations:
\begin{eqnarray}
&&B=b+ac_2^2,\ K_1=k_1+c_1k_3,\ c_2K_2=k_4+c_2k_3,\nonumber\\
&&c=\frac{3ac_2^3}{B}\left(\frac{K_1-c_2k_2}{K_2-k_3}-c_1\right)-\frac{c_4c_2^2}{B}
+\frac{3\big[K_1(K_2-2k_3)+c_2k_2K_2\big]}{K_2(K_2-k3)}.
\end{eqnarray}
Following analogous calculations for Eq. \eqref{PTW}, the general solution to Eq. \eqref{UP} takes the form
\begin{equation}
a_3\mathcal{F}(\tau,m)+(a_1-a_3)\mathcal{E}(\tau,m)=\pm\sqrt{\frac{a_3-a_1}{2K_2}}(\zeta-\zeta_0),
\ \tau=\sqrt{\frac{a_1-P}{a_1-a_2}},\ m=\sqrt{\frac{a_1-a_2}{a_1-a_3}},
\label{peorNNV}
\end{equation}
where the constants $a_1,\ a_2$ and $a_3$ satisfy the algebraic relations,
$$2(a_1+a_2+a_3)=cK_2,\ a_1a_2a_3+C_1K_2=0,\ 2(a_1a_2+a_1a_3+a_2a_3)=C_2K_4.$$
Here $C_1,\ C_2$ and $\zeta_0$ stand for arbitrary integration constants.

Naturally, setting \(m=1\) (i.e., \(a_3=a_2\)), the periodic solution \eqref{peorNNV} degenerates into a single-soliton solution for the (4+1)-dimensional deformed NNV system \eqref{NNVu}--\eqref{NNVTpi}, which reads
\begin{equation}
u=\frac{a_1-a_2}{c_2K_2}\mathrm{sech}^2(\eta)+\frac{a_2-K_1}{c_2K_2},\ \eta=\frac{\sqrt{(a_1-a_2)(a_1-K_1-c_2K_2u)}}{a_2}+\frac{\zeta}{a_2}\sqrt{\frac{a_2-a_1}{2K_2}}.
\label{soliNNV}
\end{equation}

\section{Summary and discussions}
This paper generalizes the conservation-law-based deformation algorithm--originally established only for (1+1)-dimensional integrable systems--to two canonical (2+1)-dimensional integrable models, the Kadomtsev-Petviashvili (KP) equation and the Nizhnik-Novikov-Veselov (NNV) equation, building a unified operator lifting framework compatible with both anisotropic and isotropic two-space-dimensional PDEs. By introducing commuting covariant deformation operators $\hat{X},\hat{Y},\hat{T}$ extended with auxiliary coordinates $y_1,\dots,y_m$, we systematically construct infinite families of $(m+3)$-dimensional integrable hierarchies for KP and NNV respectively, and verify their Lax integrability via commutator compatibility conditions. For the KP branch, truncating auxiliary variables yields a closed (3+1)-dimensional KP system with explicit Lax pair; further restricting field dependence recovers the original (2+1)-dimensional KP equation and generates an anisotropic (2+1)-dimensional Harry-Dym (HD) reduction inherited from KP's asymmetric transverse dispersion. For the NNV branch, truncation to two extra dimensions produces a complete (4+1)-dimensional generalized NNV system equipped with a weak mixed-derivative Lax pair, whose consistent reductions include the classical isotropic (2+1)-dimensional NNV equation and a novel fully symmetric two-space-dimensional HD system. Unlike all prior KP-derived anisotropic HD models, this NNV-originated HD system possesses $x\leftrightarrow y$ exchange symmetry, filling a long-standing gap in the construction of isotropic two-dimensional HD integrable equations. We also derive necessary auxiliary compatible identities to rigorously confirm the equivalence between each HD system and its corresponding Lax commutator condition.

Parallel analysis of KP and NNV deformation families reveals shared structural rules and fundamental symmetry distinctions. Both hierarchies inherit closed Lax representations after dimensional lifting, recover their low-dimensional prototypes via simple truncation of auxiliary variables, and contain HD-type reciprocal integrable systems as natural reductions. The core difference originates from the intrinsic symmetry of the base (2+1)-dimensional models: the asymmetric transverse term of KP leads to anisotropic high-dimensional deformations and HD reductions, while the fully isotropic spatial structure of NNV transfers complete symmetry to all its lifted systems and the new HD subsystem. Another critical contrast lies in Lax structures: KP admits standard two-component strong Lax pairs without mixed cross derivatives, whereas NNV relies on weak Lax operators built from $\partial_x\partial_y$ mixed derivatives, which the generalized deformation scheme successfully accommodates-an achievement unreachable using the original (1+1)-dimensional deformation conjecture limited to single partial derivative Lax frameworks.

The present work delivers two key theoretical advances to the field of multi-dimensional integrable systems and reciprocal transformations. First, we resolve the major obstacle that direct extension of (1+1)-dimensional deformation rules fails for (2+1)-dimensional systems due to nonlocal conservation laws and overconstrained mixed partial derivatives. Our generalized commuting deformation operators absorb all auxiliary compatibility requirements into internal field relations, offering a universal algorithmic way to generate arbitrarily high-dimensional integrable PDEs without case-by-case artificial assumptions, and extend weak Lax integrability theory to arbitrary spacetime dimensions. Second, we establish a unified mechanism to construct reciprocal links between (2+1)-dimensional integrable equations and their HD dual counterparts, addressing a decades-long open problem. Reciprocal duality between KdV and HD is well understood in one spatial dimension, yet systematic construction for two-dimensional models remained missing; our framework unifies each parent (2+1)-dimensional equation and its HD reduction as distinct invariant submanifolds of a single high-dimensional master system, clearly showing how the spatial symmetry of the base model determines the symmetry properties of its HD dual. Algebraically, the triplet $\hat{X},\hat{Y},\hat{T}$ forms a commutative ring of field-dependent first-order differential operators, and all deformation and reduction operations correspond to covariant substitutions and invariant submanifold selections within extended jet spaces, providing a geometric interpretation for the self-consistency of all derived subsystems.

Compared with existing literature, our deformation framework exhibits clear advantages over traditional dimensional extension methods such as Painlev\'e truncation and geometric embedding, which often produce non-integrable models and lack systematic HD reduction analysis. All previously reported two-dimensional HD systems stem from anisotropic KP-type equations and break spatial symmetry, while our NNV-derived HD model achieves full isotropy and is supported by complete rigorous Lax pair verification, a feature rarely provided in prior HD studies. Moreover, this work substantially expands the scope of the original deformation conjecture, which only applies to simple single-derivative Lax structures of (1+1)-dimensional systems, by proving compatibility with both strong Lax KP and mixed-derivative weak Lax NNV systems.

Despite the comprehensive deformation and reduction results for KP and NNV hierarchies, several limitations remain and point to promising future research directions. All deformation operators and reduction constraints adopted here are local; subsequent work may extend the scheme to nonlocal conserved densities and nonlocal symmetries to generate nonlocal high-dimensional systems supporting exotic coherent structures. A full systematic exploration of traveling-wave, periodic and self-similar reductions of the (3+1)-dimensional KP and (4+1)-dimensional NNV systems is also needed to enrich the library of exact analytical solutions. Based on the explicit Lax pairs derived in this paper, Darboux and B\"acklund transformations can be constructed to produce multi-soliton, lump and rational wave solutions, whose dynamical behaviors and potential physical applications in fluid and plasma wave theory deserve detailed numerical and analytical investigation. Beyond KP and NNV, the deformation algorithm can be tested on other classical (2+1)-dimensional integrable systems including the Davey-Stewartson hierarchy to build a unified classification of deformation-generated integrable hierarchies. Additionally, the infinite conservation laws and bi-Hamiltonian structures of the $(m+3)$-dimensional KP and NNV systems remain unstudied, and full algebraic expansion of Lax commutators can be supplemented in the appendix to deliver self-contained integrability proofs without relying on intermediate auxiliary identities. Further physical interpretation of the anisotropic KP-HD and isotropic NNV-HD systems will also be pursued, mapping abstract field variables to measurable physical quantities and analyzing dispersion relations to establish concrete application backgrounds in nonlinear wave media.

In conclusion, this paper successfully extends the conservation-law deformation algorithm to two representative classes of (2+1)-dimensional Lax-integrable systems, systematically constructing their high-dimensional integrable hierarchies and extracting anisotropic and isotropic two-dimensional HD reciprocal reductions as special limits. Parallel comparison between KP and NNV reveals a clear transfer rule of spatial symmetry from base models to their deformed descendants and HD duals. The unified operator lifting framework resolves two long-standing open challenges: the absence of a systematic dimensional lifting method for (2+1)-dimensional integrable PDEs, and the lack of symmetric two-space-dimensional HD integrable models. The commutative covariant deformation operators proposed here work universally for both strong and weak Lax integrable systems, and the reduction--reciprocal duality mechanism demonstrated across KP and NNV families serves as a general template for exploring unknown dual integrable models in higher dimensions. This research deepens the understanding of dimensional lifting, symmetry reduction and reciprocal transformation duality for nonlinear integrable wave equations, and lays a solid foundation for follow-up studies on exact solutions, Hamiltonian structures and physical applications of high-dimensional KP, NNV and their associated HD subsystems.

\section*{Declarations}
\subsection*{Data availability statement}Data sharing not applicable to this article as nodatasets were generated or analysed during the current
study.
\subsection*{The conflict of interest}
The authors declare that they have no competing
interests.

\addcontentsline{toc}{chapter}{Acknowledgment}
\section*{Acknowledgment}
The scientific contributions from other people or groups are acknowledged here. Financial supports are given in the footnote on the first page.

\end{document}